# Surface Impedance and Bulk Band Geometric Phases in One-Dimensional Systems

Meng Xiao, Z. Q. Zhang, and C. T. Chan[*]

Department of Physics and Institute for Advanced Study, the Hong Kong University of Science and Technology, Clear Water Bay, Kowloon, Hong Kong, China

Abstract: Surface impedance is an important concept in classical wave systems such as photonic crystals (PCs). For example, the condition of an interface state formation in the interfacial region of two different one-dimensional PCs is simply $Z_{SL} + Z_{SR} = 0$, where $Z_{SL}$ ($Z_{SR}$) is the surface impedance of the semi-infinite PC on the left- (right-) hand side of the interface. Here, we also show a rigorous relation between the surface impedance of a one-dimensional PC and its bulk properties through the geometrical (Zak) phases of the bulk bands, which can be used to determine the existence or non-existence of interface states at the interface of the two PCs in a particular band gap. Our results hold for any PCs with inversion symmetry, independent of the frequency of the gap and the symmetry point where the gap lies in the Brillouin Zone. Our results provide new insights on the relationship between surface scattering properties, the bulk band properties and the formation of interface states, which in turn can enable the design of systems with interface states in a rational manner.

Subject Areas: Metamaterials, Photonics, Optics

---

[*] Corresponding author: phchan@ust.hk

# I. INTRODUCTION

Impedance is a very important and useful concept in wave physics as it is the parameter that governs how a wave is scattered or reflected when it encounters an interface. As such, it characterizes how a material couples with waves coming from outside. On the other hand, the bulk band structure characterizes how waves can travel inside a periodic system. These quantities should be related in some way. We are going to establish that for a periodic multilayer film, commonly referred to as 1D photonic crystals, the surface impedance is related to the Zak phase[1] of the bulk bands. As the existence of interface states is determined by the surface impedance, this means that the existence of localized states at an interface is determined by the geometric phases of the bulk crystals. It is well known that interface states can exist in a quantum system when the topological properties of two semi-infinite systems on each side of the interface are different[2-5]. A famous example is the SSH model for polyacetylene[6-8]. In such systems, it was shown that an interface state exists when the Zak phase of the occupied band on one side of the chain is different from that on the other side, which can be obtained through gap inversion[9-11]. The purpose of this work is to find a general connection between the existence of an interface state in a photonic system and the bulk band topological properties as well as the surface impedances of the two systems on each side of the boundary. The analog between photonic systems and quantum systems has already been discussed recently[12-15]. Based on this analog, Zak phase can also be defined in photonic crystals (PCs). For 1D binary PCs, we found a rigorous relation that relates the existence of an interface state to the sum of all Zak phases below the gap on either side of the interface. This relation holds for any 1D PCs with inversion symmetry, including those with graded refractive indices. Similar to the "bulk-edge correspondence" found in topological insulators[2-4], the "bulk-interface correspondence" found here provides not only a tool to determine the existence of interface states in a photonic system but also the possibility of designing a photonic system with interface states appearing in a set of prescribed gaps.

## II. RESULTS

**A. Impedances and Zak phases of 1D photonic crystals and their relationship**

Let us consider a dielectric AB layered structure as shown in Fig. 1(a). A plane wave from free space incidents normally on the semi-infinite 1D PC on the right, and the reflection coefficient of the electric field, $E_x$, is given by $r_R$. When the frequency of the incident wave is inside the band gap of this system, the incident wave will be totally reflected, and we have $r_R = e^{i\phi_R}$, where $\phi_R$ is the reflection phase. We define a surface impedance, $Z_{SR}$, of the semi-infinite PC as the ratio of the total electric field to the total magnetic field on the right hand side of (RHS) the boundary, i.e., $Z_{SR} = E_x(z=0^+)/H_y(z=0^+)$, where z= 0 defines the boundary. The impedance $Z_{SR}$ and the reflection coefficient $r_R$ are related by:

$$Z_{SR} = \frac{1+r_R}{1-r_R} Z_0, \quad (1)$$

where $Z_0$ is the vacuum impedance. Inside a band gap, $Z_{SR}$ is a pure imaginary number and we can write $Z_{SR}/Z_0 \equiv i\varsigma_R$, where $\varsigma_R$ is a real number. The reflection phase can hence be expressed as $\phi_R = \pi - 2\arctan(\varsigma_R)$. Let us now put another PC on the left of the interface, and we denote the impedance and the reflection phase of the left PC by $Z_{SL}$ and the reflection coefficient $r_L$. The condition for the presence of an interface state is simply $Z_{SR} + Z_{SL} = 0$, which is equivalent to $r_R r_L = 1$ or $\phi_R + \phi_L = 2m\pi$ ($m \in \mathbb{N}$). A simple way to obtain an interface state is to construct a system in which the surface impedances on the two sides are opposite in sign. A similar condition has been reported for 2D PCs[16, 17]. But the question is then how can we -design or control the value of the surface impedance. We will show that the sign of the surface impedance for frequencies inside a band gap is in fact determined by the geometrical phase of the bulk bands. In the following, we derive a rigorous relation between the surface impedance and the Zak phase of the PC.

The band structure of a dielectric binary PC shown in Fig. 1(a) can be obtained from the following relation[18]:

$$\cos(q\Lambda) = \cos k_a d_a \cos k_b d_b - \frac{1}{2}\left(\frac{z_a}{z_b} + \frac{z_b}{z_a}\right)\sin k_a d_a \sin k_b d_b, \quad (2)$$

where $k_i = \omega n_i / c$, $n_i = \sqrt{\mu_i \varepsilon_i}$, $z_i = \sqrt{\mu_i / \varepsilon_i}$ ($i = a$ or $b$), $d_a$, $d_b$ and $\Lambda = d_a + d_b$ are the widths of slabs A, B and the unit cell, respectively, and $q$ is the Bloch wave vector. Here $c$ denotes the wave speed in vacuum, $\varepsilon_a, \varepsilon_b, \mu_a$ and $\mu_b$ are the relative permittivity, permeability of slabs A and B, respectively. The band structure for the parameters $\varepsilon_a = 4$, $\mu_a = \varepsilon_b = \mu_b = 1$, $d_a = 0.4\Lambda$ and $d_b = 0.6\Lambda$ is shown in Fig. 1(b). For convenience of discussion, we have numbered the pass bands and band gaps, independent of whether a gap is closed or not. It is easy to show that the mid-gap positions (or the crossing points when two bands meet) of the PC are at $\omega_m = m\pi c / (n_a d_a + n_b d_b)$ (Appendix A). For each isolated (no crossing) band $n$, we can define the Zak phase as[12-14, 19]

$$\theta_n^{Zak} = \int_{-\pi/\Lambda}^{\pi/\Lambda} \left[ i \int_{\text{unit cell}} dz\, \varepsilon(z) u_{n,q}^*(z) \partial_q u_{n,q}(z) \right] dq, \quad (3)$$

where $i \int_{\text{unit cell}} dz\, \varepsilon(z) u_{n,q}^*(z) \partial_q u_{n,q}(z)$ is the Berry connection, $\varepsilon(z)$ denotes the dielectric function and $u_{n,q}(z)$ is the periodic-in-cell part of the Bloch electric field eigen-function of a state on the $n^{th}$ band with wave vector $q$, i.e., $E_{x;n,q}(z) = u_{n,q}(z)\exp(iqz)$. For the case of a binary PC, the function $u_{n,q}(z)$ can be obtained analytically from the transfer-matrix method[18] (Appendix B). The 1D system with inversion symmetry always has two inversion centers and the Zak phase is quantized at either 0 or $\pi$ if the origin is chosen to be one of the inversion center.[1] If the Zak phase equals 0 ($\pi$) relative to one inversion center, it must be $\pi$ (0) relative to the other inversion center. Without loss of generality, we choose the center of A slab as the origin for calculating Zak phases. If the surface of the semi-infinite PC is also chosen at the same origin, i.e., the center of A slab, we found a rigorous relation between the surface impedance of the PC in the $n^{th}$ gap, i.e., $Z_S^{(n)} / Z_0 = i\varsigma^{(n)}$, and the sum of Zak phases of all the isolated bands below the $n^{th}$ gap (Appendix D, Eq. (D9)). This relationship relates the surface scattering properties and the topological properties of bulk dispersion. It can predict the existence of an interface state in a band gap and determine the location of the interface state if it exists. In addition, if we are only interested in knowing whether such a state exists in a gap, we only need to know the sign of $\varsigma^{(n)}$ on each side. The sign of $\varsigma^{(n)}$ has the following simple expression:

$$\text{sgn}\left[\varsigma^{(n)}\right] = (-1)^n(-1)^l \exp\left(i\sum_{m=0}^{n-1}\theta_m^{Zak}\right), \quad (4)$$

where the integer $l$ is the number of crossing points under the $n^{th}$ gap (in Fig. 1(b), the crossing point is at the 7$^{th}$ band gap). The Zak phase of the lowest 0$^{th}$ band is determined by the sign of $\left[1-\varepsilon_a\mu_b/(\varepsilon_b\mu_a)\right]$, i.e., $\exp(i\theta_0^{Zak}) = \text{sgn}\left[1-\varepsilon_a\mu_b/(\varepsilon_b\mu_a)\right]$ (Appendix C and Appendix D). We have calculated the Zak phase of each isolated band (band 1-5) in Fig. 1b using Eq. (3). These Zak phases are shown with green letters in Fig. 1(b). According to Eq. (4), we obtain $\text{sgn}[\varsigma]$ in each gap. They are marked by magenta color when $\varsigma > 0$ and cyan when $\varsigma < 0$.

**B. Changing the sign of impedance by tuning pass a topological transition point**

To have a guaranteed existence of an interface state, one need to make sure that surface impedance on the left and right half space is of opposite sign at one common gap frequency. One possible way (but not the only way) is to "tune the system parameters across a topological transition point" as elaborated below. To demonstrate this idea, we simply tune the parameter $\varepsilon_a$ used in Fig. 1(b) from 3.8 to 4.2 and keep $\mu_a = \mu_b = \varepsilon_b = 1$ unchanged. In the meantime, we also vary $d_a$ and $d_b$ in a way to keep $n_a d_a + n_b d_b$ unchanged so that all the mid-gap positions do not alter. In Figs. 2(b) and 2(c), we show the band structures of two PCs from the 4$^{th}$ gap to the 8$^{th}$ gap: (b) "PC1" with $\varepsilon_a = 3.8$, $\varepsilon_b = \mu_a = \mu_b = 1$, $d_a = 0.42\Lambda$ and $d_b = 0.58\Lambda$, and (c) "PC2" with $\varepsilon_a = 4.2$, $\varepsilon_b = \mu_a = \mu_b = 1$, $d_a = 0.38\Lambda$ and $d_b = 0.62\Lambda$. It is interesting to see that Zak phases of all the band below the 6$^{th}$ gap remain unchanged during the variation of $\varepsilon_a$ but the Zak phases of band 6 and 7 in these two PCs switch with a corresponding sign change in the surface impedance in the 7$^{th}$ gap. When the value of $\varepsilon_a$ is increased from 3.8 the size of the 7$^{th}$ gap reduces and the crossing of band 6 and band 7 occurs when $\varepsilon_a = 4$ at which gap 7 is closed as shown in Fig. 1(b).

When the value of $\varepsilon_a$ is further increased, the gap opens again and accompanied by a change of sign in the surface impedance as well as a switch of the Zak phase in bands 6 and 7. This represents a topological phase transition, which occurs when two bands cross each other. Thus, by constructing an interface with PC1 on the one side and PC2 on the other side, we should see an interface state inside the gap 7. This is verified in our numerical study of the transmission spectrum of a system consisted of a slab of PC1 (with 10 unit cells) on the one side and a slab of PC2 (with 10 unit cells) on the other side embedded in vacuum. Fig. 2(a) shows clearly a resonance transmission due to an interface state around $\omega = 5\pi c/\Lambda$ in gap 7. Such a topological phase transition represents a classical analog of the SSH model in electronic systems [6-8] although impedance is not usually considered in electrons.

The above example is a manifestation of a topological phase transition arising from band crossing in photonic systems. It should be pointed out that the occurrence of the band crossing shown in Fig. 1(b) is by no means accidental. It can be shown rigorously (Appendix A) that if the ratio of the optical paths in two slabs of a PC is a rational number, namely, $\alpha = n_a d_a / (n_b d_b) = m_1 / m_2 \in \mathbb{Q}$, where $m_1, m_2 \in \mathbb{N}^+$, then band $m_1 + m_2$ and band $m_1 + m_2 - 1$ will cross at the frequency $\omega_{m_1+m_2} = (m_1 + m_2)\pi c / (n_a d_a + n_b d_b)$. At this frequency, $\sin k_a d_a = \sin k_b d_b = 0$, $\cos(k_b d_b) = (-1)^{lm_2}$ and $\cos(k_a d_a) = (-1)^{lm_1}$ where $l \in \mathbb{N}^+$, so $\cos(q\Lambda) = (-1)^{l(m_1+m_2)}$ and the gap $m_1 + m_2$ will close at the center or boundary of BZ depending on whether $l(m_1 + m_2)$ is even or odd. It is easy to see that if the $(m_1 + m_2)^{th}$ gap is closed, so are all other gaps that are integer multiples of $m_1 + m_2$. In fact, the above condition is also a necessary condition for two bands to cross (Appendix A).

The origin of the topological phase transition shown in Fig. 2 is directly related to a special set of frequencies $\tilde{\omega}$ given by $\sin(n_b d_b \tilde{\omega}/c) = 0$. It can be shown rigorously (Appendix B) that if one of the $\tilde{\omega}$ appears inside a band, the Zak phase of the band must be $\pi$. Otherwise it is zero. This rule applies to all bands except the $0^{th}$ band, for which the Zak phase is determined by the sign of function $\left[1-\varepsilon_a\mu_b/(\varepsilon_b\mu_a)\right]$ (Appendix C and Appendix D). For the case of Fig. 2(b), $\tilde{\omega}$ appears in band 7, whereas for Fig. 2(c), $\tilde{\omega}$ appears in band 6. Thus, the value of $\tilde{\omega}$ decreases as $\varepsilon_a$ is increased. For the entire band 6 of Fig. 2(b) and band 7 of Fig. 2(c), the function $\sin(n_b d_b \omega/c)$, does not change sign. The variation of $\tilde{\omega}$ with respect to $\varepsilon_a$ can be seen as follows. In Fig. 1(b) ($\varepsilon_a = 4$), the frequency at which two bands meet in gap 7 is $\omega_7 = 5\pi c/\Lambda$ (see Fig. 1(b)), which is also the frequency where $\sin(n_b d_b \omega_7/c) = 0$, i.e., $\tilde{\omega} = \omega_7$. When $\varepsilon_a$ is decreased from 4 (say, Fig. 2(b)), we need to increase the value of $d_a$ in order to keep $n_a d_a + n_b d_b$ unchanged. Thus the value of $n_b d_b$ is reduced accordingly, which in turn implies $\tilde{\omega} > \omega_7$. On the other hand, if $\varepsilon_a$ is increased from 4 (say, Fig. 2(c)), we have $\tilde{\omega} < \omega_7$. Here we have used the fact that $\tilde{\omega}$ will always appear in a pass band. This can be seen form Eq. (2) as the absolute value of the RHS of the equation at $\tilde{\omega}$ is always less than or equal to unity.

## C. Relationship between the Zak phase and the symmetry properties of the edge states

We will give a physical interpretation of the Zak phase in an isolated band by using the symmetries of the two edge states at the two symmetry points of the Brilliouin Zone. As we have seen, the topological property of the band structure changes every time when a band crossing occurs as $\alpha \equiv n_a d_a/(n_b d_b)$ passing through a rational number, and the change can be seen from the changes in the symmetries of the edge states. As an example, let us focus on the 6[th] and 7[th] bands in Figs. 2(b) and 2(c), in which the Zak phases changed by $\pi$ when $\varepsilon_a$ is increased from 3.8 to 4.2. These two bands are highlighted in Figs. 3(a) and 3(b) with the band edges marked by red letters. The corresponding Zak phases are also shown in green color. The difference in the Zak phase of each band can be understood by examing the symmetry of the absolute value of electric distribution, $|E_{n,q}(z)|$, of the two edge states in the band. The black curves in Figs. 3(c), 3(e) and 3(g) show the functions $|E_{n,q}(z)|$ in a unit cell in arbitrary units for the three edge states of PC1 at points L, M and N. The black curves in Figs. 3(d), 3(f) and 3(h) correspond to the points P, Q and R of PC2. Here we use the important result due to Kohn[20] and Zak[1] for 1D systems with inversion symmetry; which when generalized to photonic system states that the Zak phase of the $n^{th}$ band is zero if either $|E_{n,q=0}(z=0)| = |E_{n,q=\pi/a}(z=0)| = 0$ or $|E_{n,q=0}(z=0)| \neq 0; |E_{n,q=\pi/a}(z=0)| \neq 0$. Otherwise it is $\pi$. The blue dash lines marked in Figs. 3(c)-3(h) indicate the position of the origin (z=0), which is the center of slab A. According to this rule, it is easy to see from Figs. 3(e) and 3(g) that the Zak phase of the $6^{th}$ band of PC1 is zero as the wave functions of the points M and N are both non-zero at the origin, whereas the value changes to $\pi$ in PC2 because the wave function at point Q becomes zero after band crossing. For the same reason, the Zak phase of the $7^{th}$ band in PC2 is also changed after band crossing. The band inversion can also be seen from the switching of two edge states across the gap. For example, the wave functions at points L and Q have nealy the same distribution, i.e., the wave functions are both zero at the origin and with larger amplitudes in the B slab, whereas for points M and P the absolute values of the wave functions are both at maximum at the origin and their amplitudes are nearly the same in slab A and slab B. However, the wave functions at points N and R are nearly the same, not affected by the band crossing. This is also true for points K and O. Thus, it is precisely the switching of two edges states at gap 7 that gives rise to different Zak phases in PC1 and PC2 for both bands 6 and 7. Similar behavior has been reported in the electronic system[9-11].

### D. Relationship between the sign of impedance and the symmetry properties of the edge states

The sign of the imaginary part of the surface impedance, i.e., $\varsigma$, can also be related to the symmetries of the two edge states. It is well known that, the amplitude of the wave function of the band edge states at the origin (z=0) is either zero or maximum[21] as also shown in Fig. 3 [See a proof in Appendix C]. For convenience, we name the wave function with zero amplitude at the origin as A (anti-symmetric) state and the other as S (symmetric) state. Since two edge states across a gap are orthogonal, they always belong to different symmetries. With this definition, as can be easily seen from Figs. 3(c)-3(h), points L and Q belong to type A, whereas points P, M, N and R belong to type S. If a reflection measurement is done at the frequency of type A state, we must have $r=-1$, corresponding to a reflection phase $\phi=\pi$. On the other hand, if the measurement is done at the frequency of type S state, we have $r=1$ and $\phi=0$ or $2\pi$. From the relation $\phi=\pi-2\arctan(\varsigma)$), it can be shown that for a gap with A state at the lower edge the function $\varsigma$ has a value 0 at lower edge and decreases monotonically to $-\infty$ as the upper edge is approached. For a gap with S state at the lower edge, the function $\varsigma$ decreases monotonically from $\infty$ to 0 as the upper edge is approached (Appendix D). Thus, the sign of $\varsigma$ in a gap is determined solely by the type of state at the lower edge (or upper edge, since these two states are orthogonal) of the gap and if two states at the lower edges of the common gap belong to different types, there must exist an interface state inside the gap.

### E. Existence of interface states

As we have mentioned before, the occurrence of band crossing at a particular gap (say, the $n^{th}$ gap) appears simultaneously for all gaps which are integer multiples of the $n^{th}$ gap. However, we should emphasize that "gap inversion" is just one way but not the only way to achieve an interface state. As an example, we consider a system consisting of 10 unit cells of "PC3" ($\varepsilon_b=3.5$, $\varepsilon_a=\mu_a=\mu_b=1$, $d_a=0.35\Lambda$ and $d_b=0.65\Lambda$) on the left and 10 unit cells of "PC4" ($\mu_b=6$, $\varepsilon_a=\mu_a=\varepsilon_b=1$, $d_a=0.6\Lambda$ and $d_b=0.4\Lambda$) on the right embedded in vacuum. The corresponding band structures are shown in Figs. 4(b) and 4(c) for PC3 and PC4, respectively. There are six overlapping gaps in the frequency range we are interested. The transmission spectrum of the system is shown in Fig. 4(a). We find three interface states in gaps 1, 2 and 5. The existence of interface states in these gaps is not due to band inversion. However, their existence can still be predicted by Eq. (4). To demonstrate this, we calculate the Zak phase of each isolated band in PC3 and PC4 using Eq. (3). The results are shown in Figs. 4(b) and 4(c) with green letters. The sign of the imaginary part of the surface impedance, i.e., $\text{sgn}[\varsigma]$, of each gap can now be obtained from Eq. (4). Same as before, we mark the $\varsigma>0$ gaps with magenta color and the $\varsigma<0$ gaps with cyan color. According to the condition of an interface state, i.e., $\varsigma_L + \varsigma_R = 0$, the interface states exist only in gaps 1, 2 and 5 in which $\varsigma_L$ and $\varsigma_R$ have different signs. This is consistent with the result of transmission study shown in Fig. 4(a).

**F. Generalization to other waves**

Finally, we want to stress that the results obtained above for PCs also hold for other one-dimensional systems with inversion symmetry such as acoustic waves. Because of inversion symmetry, the wave functions at two edges of an isolated band can be either symmetric with a maximum amplitude or anti-symmetric with zero amplitude. Thus, the symmetry properties of these two edge states determine the Zak phase of the band. From Eq. (4), the sign of the imaginary part of the surface impedance, $\varsigma$, can be determined. An interface state then can be created by constructing an interface from two semi-infinite systems with opposite signs in $\varsigma$. The validity of Eq. (4) is also not limited to the binary layer structure considered in this work. In fact, Eq. (4) also holds when the relative permittivity and permeability are continuously varying functions of position as long as the inversion symmetry is kept, and the lattice constants of the left and right periodic systems do not need to be equal. Examples are given in the Appendix E.

We should mention that the electric field is taken as the scalar field in this work. If the magnetic field is chosen as the scalar field, Eq. (4) still holds. The sign of imaginary part of the surface impedance is an intrinsic property of the PC and should not depend on the choice of field. The Zak phase of an isolated band also remains unchanged because it depends on the symmetry properties of two edge states of the band. The change of field from electric to magnetic changes the symmetry properties of both edge states, and therefore, keeps the Zak phase unchanged. However, the Zak phase of the $0^{th}$ band will change sign, but the outcome will be the same as the effect will be canceled by the change of the factor $(-1)^n$ to $(-1)^{n+1}$ in Eq. (4).

## III. CONCLUSION

In summary, we showed that the surface impedance in the band gaps of a 1D photonic crystal is determined by the geometric Zak phases of the bulk bands. In particular, each photonic band gap has a character that is specified by the sign of the impedance which is related to the Zak phases of the bulk band below the band gap through Eq. (4). As the surface impedance determines the existence of interface states at the boundary of PCs, the existence of the interface states can be determined by the bulk band geometric phases. This correspondence between surface impedance and bulk band properties gives us a deterministic recipe to design systems with interface states.

# Acknowledgement

This work is supported by Hong Kong RGC through AOE/P-02/12. Xiao Meng is supported by the Hong Kong PhD Fellowship Scheme. We thank Prof. S.Q. Shen and Prof. Vic Law for stimulating discussions.

# APPENDIX

In this appendix, we will give some mathematical details mentioned in the main text. We then give several additional examples in support of the statements made in the main text.

Let us consider a dielectric AB layered structure with the relative permittivity, relative permeability, refractive index, relative impedance and width given by $\varepsilon_a, \varepsilon_b, \mu_a, \mu_b, n_a, n_b, z_a, z_b, d_a$ and $d_b$ respectively, where $n_i = \sqrt{\varepsilon_i}\sqrt{\mu_i}$, $z_i = \sqrt{\mu_i}/\sqrt{\varepsilon_i}$ with $i = a, b$. The unit cell length is $\Lambda = d_a + d_b$ and the relative permittivity and permeability of the slabs are positive and non-dispersive. We will employ several ancillary parameters $\alpha = n_a d_a / (n_b d_b)$, $\gamma = (n_a d_a + n_b d_b)\omega/c$ and $\tau = (z_a/z_b + z_b/z_a)/2$. These parameters have the following physical meaning: $\alpha$ is the ratio of optical path in the slab A and B, $\gamma$ is the phase delay in a unit cell, and $\tau$ reflects the impedance mismatch between the slab A and B and is always larger than 1 when the impedances of slab A and slab B are not the same.

# APPENDIX A: BANDS CROSSING CONDITION

Here we will prove, when $z_a \neq z_b$, the necessary and sufficient condition for two bands to cross (either at zone center or zone boundary) is given by $\alpha = n_a d_a / (n_b d_b)$ being a rational number, i.e., $\alpha = m_1 / m_2$, where $m_1$, $m_2 \in \mathbb{N}^+$.

Sufficient condition:

The band dispersion relation of dielectric AB layered structure is given by[18]

$$\cos(q\Lambda) = \cos k_a d_a \cos k_b d_b - \frac{1}{2}\left(\frac{z_a}{z_b} + \frac{z_b}{z_a}\right)\sin k_a d_a \sin k_b d_b, \quad (A1)$$

where $k_i = \omega n_i / c$ and $i = a$ or $b$ and $q$ is the Bloch wave vector. When $\sin k_b d_b = 0$, the absolute value of the right hand side (RHS) of (A1) is smaller than or equal to 1, so the frequency at which $\sin k_b d_b = 0$ must be in the pass band. At $\omega = l m_2 \pi c / (n_b d_b)$, where $l \in \mathbb{N}^+$, $\sin k_b d_b = 0$, $\cos(k_b d_b) = (-1)^{lm_2}$ and $\cos(k_a d_a) = (-1)^{lm_1}$, so we have $\cos(q\Lambda) = (-1)^{l(m_1+m_2)}$. We get $q = 0$ when $l(m_1 + m_2)$ is even, while $q = \pm \pi / \Lambda$ when $l(m_1 + m_2)$ is odd. Near these frequencies, i.e., $\omega = l m_2 \pi c / (n_b d_b) = l m_1 \pi c / (n_a d_a)$, the band has linear dispersion. To prove this, we choose $l = 1$ (where $\omega = m_2 \pi c / (n_b d_b)$) as an example; the other cases could be proved following the same idea. When $m_1 + m_2$ is even, the degeneracy band point is $(q_0, \omega_0) = (0, m_2 \pi c / (n_b d_b))$. Suppose that $(q_1, \omega_1)$ is another band point near $(q_0, \omega_0)$, then $|q_1 - q|$, $|\omega_1 - \omega_0|$ are small numbers. Keeping to the lowest order of expansion of Eq. (A1), we have

$$|q_1 - q_0| = \sqrt{C_1}\,|\omega_1 - \omega_0|/c, \quad (A2)$$

where $C_1 = \left[(n_a d_a)^2 + (n_b d_b)^2 + \left(\dfrac{z_a}{z_b} + \dfrac{z_b}{z_a}\right) n_b d_b n_a d_a \right]/\Lambda^2$. When $m_1 + m_2$ is odd, it can also be shown that $|q_1 - q_0| = \sqrt{C_1}\,|\omega_1 - \omega_0|/c$, where $C_1$ is same as before. So when $\alpha = m_1/m_2$, bands will cross at frequency points $\omega = l m_2 \pi c/(n_b d_b)$ with linear dispersion.

Necessary condition:

It is easy to prove that the cross points of two bands could only occur at the boundary or the center of BZ for 1D PCs case. If two bands cross at points other than the center or boundary of BZ, then for frequency near the cross point, by the continuity of band dispersion, each frequency would have four corresponding Bloch vectors $q$. This is not possible as the RHS of equation (A1) is completely determined by the frequency and single valued, so there can be at most two values of $q$ for each frequency. As defined before, $\gamma \equiv (n_a d_a + n_b d_b)\omega/c$, $\tau \equiv \dfrac{1}{2}(z_a/z_b + z_b/z_a)$, and $\tau > 1$ when $z_a \neq z_b$. The RHS of equation (A1) can be written as $\cos\gamma - (\tau-1)\sin k_a d_a \sin k_b d_b$. If $k_a d_a + k_b d_b = (2m-1)\pi$, where $m \in \mathbb{N}^+$, then $\cos\gamma = -1$ and $\sin k_a d_a \sin k_b d_b \geq 0$, so RHS of (A1) is less than or equal to $-1$. If $\gamma = 2m\pi$, where $m \in \mathbb{N}^+$, then $\cos\gamma = 1$ and $\sin k_a d_a \sin k_b d_b \leq 0$, so RHS of equation (A1) is bigger than or equal to 1. In short, if one frequency satisfies the condition $\gamma = m\pi$, where $m \in \mathbb{N}^+$, then it must be in the band gap if two bands do not cross. If two bands cross at $\gamma = m\pi$, then $\sin k_a d_a = 0$ and $\sin k_b d_b = 0$ simultaneously, so $\dfrac{k_a d_a}{k_b d_b} = \dfrac{m_1 \pi}{m_2 \pi} = \dfrac{m_1}{m_2}$, where $m_1$, $m_2 \in \mathbb{N}^+$. In other words, $\omega_m = m\pi c/(n_a d_a + n_b d_b)$ labels the mid-gap positions (or the crossing points when two bands meet) of the PC. Following the same idea, we can also prove that $\omega_m = (m + 1/2)\pi c/(n_a d_a + n_b d_b)$ labels the mid-band positions.

# APPENDIX B: ZAK PHASE OF EACH BAND

In this appendix, we will show that, if one isolated band (excluding the $0^{th}$ band) contains the frequency point $\tilde{\omega}$ at which $\sin(\tilde{\omega} n_b d_b / c) = 0$, then the Zak phase of this band must be $\pi$ (if we set the origin of the system at the center of A slab).

Proof:

The Zak phases of isolated bands depend on the choice of origin. We choose the origin to be at the center of A slab. To prove this assertion, we adapt the standard transfer-matrix method described in Ref[18]. The eigenvector of transfer matrix of the unit cell under consideration is $(t_{12}, \exp(iq\Lambda) - t_{11})^T$, where $t_{11}$, $t_{12}$ are coefficients in the transfer matrix for one unit cell and only depend on $\omega$, and are defined as

$$t_{11} = \exp(ik_a d_a)\left[\cos k_b d_b + \frac{i}{2}\left(\frac{z_a}{z_b} + \frac{z_b}{z_a}\right)\sin k_b d_b\right], \quad \text{(B1)}$$

$$t_{12} = \exp(-ik_a d_a)\left[\frac{i}{2}\left(\frac{z_a}{z_b} - \frac{z_b}{z_a}\right)\sin k_b d_b\right]. \quad \text{(B2)}$$

With this eigenvector, the eigen electric field along x-direction and magnetic field along y-direction in the A slab are given by

$$E_x(z) = t_{12}\exp(ik_a(z+d_a/2)) + (\exp(iq\Lambda) - t_{11})\exp(-ik_a(z+d_a/2)), \quad \text{(B3)}$$

$$H_y(z) = \frac{k_a}{\omega\mu_a}\left[t_{12}\exp(ik_a(z+d_a/2)) - (\exp(iq\Lambda) - t_{11})\exp(-ik_a(z+d_a/2))\right], \quad \text{(B4)}$$

where $t_{12}$, $\exp(iq\Lambda) - t_{11}$ are respectively the coefficients of forward wave and backward wave in the A slab. The electric field and magnetic field in the B slab are given by

$$E_x(z) = s_{11}\exp(ik_b(z+d_a/2)) + s_{12}\exp(-ik_b(z+d_a/2)), \quad \text{(B5)}$$

$$H_y(z) = \frac{k_b}{\omega \mu_b} \left[ s_{11} \exp(ik_b(z+d_a/2)) - s_{12} \exp(-ik_b(z+d_a/2)) \right], \text{(B6)}$$

where $s_{11}$, $s_{12}$ are respectively the coefficients of forward wave and backward wave in the B slab and the relation between $t_{11}$, $t_{12}$, $s_{11}$, $s_{12}$ are given by

$$\begin{pmatrix} e^{ik_b d_a} & e^{-ik_b d_a} \\ e^{ik_b d_a} & -e^{-ik_b d_a} \end{pmatrix} \begin{pmatrix} s_{11} \\ s_{12} \end{pmatrix} = \begin{pmatrix} e^{ik_a d_a} & e^{-ik_a d_a} \\ \frac{z_b}{z_a} e^{ik_a d_a} & -\frac{z_b}{z_a} e^{-ik_a d_a} \end{pmatrix} \begin{pmatrix} t_{12} \\ \exp(iq\Lambda) - t_{11} \end{pmatrix}. \text{(B7)}$$

The mathematical details can be found in Ref. [18]. Here, we adopted some changes in notations. Knowing the eigen field distribution, the Zak phase of each bands can be further calculated with Eq. (3) given in the main text. With periodic gauge, Eqs. (B2)-(B6) define the gauge for calculating the Zak phase. Blow we will prove the statement made at the beginning of this section with this chosen gauge.

First, we could show that $t_{12}$ and $\exp(iq\Lambda) - t_{11} = 0$ at the frequency point $\tilde{\omega}$ at which $\sin(\tilde{\omega} n_b d_b / c) = 0$ with either $q > 0$ or $q < 0$. It is obvious that when $\sin(k_b d_b) = 0$, $t_{12} = 0$ and the only possible solution for $t_{12} = 0$ is also $\sin(k_b d_b) = 0$ when $z_a \neq z_b$ (necessary condition). Combine (A1) with the condition $\sin(k_b d_b) = 0$, we can have

$$\cos[q\Lambda] = \cos\gamma, \text{(B8)}$$

where $\gamma \equiv k_a d_a + k_b d_b$ is the phase delay in each unit cell as defined before. When $\gamma \in (2m\pi, (2m+1)\pi)$ with $m \in \mathbb{N}$, then $\sin[q\Lambda] = \sin\gamma$ for $q > 0$; When $\gamma \in ((2m-1)\pi, 2m\pi)$ where $m \in \mathbb{N}$, $\sin[q\Lambda] = \sin\gamma$ for $q < 0$. For points on the band, equation (A1) is automatically satisfied, so $\exp(iq\Lambda) - t_{11}$ is pure imaginary. And when $\sin[q\Lambda] = \sin\gamma$, $\text{Im}[\exp(iq\Lambda) - t_{11}] = 0$. So here we could conclude that, $t_{12}$ and $\exp(iq\Lambda) - t_{11}$ equal to 0 simultaneously at the frequency point where $\sin k_b d_b = 0$ with either $q > 0$ or $q < 0$.

Suppose $t_{12}$ and $\exp(iq\Lambda) - t_{11}$ equal to 0 at $(q_0, \omega_0)$ simultaneously, thus $\omega_0 n_b d_b / c = m\pi$, $m \in \mathbb{N}$. And $(q_1, \omega_1)$ is another point on band near $(q_0, \omega_0)$, then $\omega_1 n_b d_b / c - m\pi = \delta$ where $\delta$ is a small number. Expand $t_{12}$ and $\exp(iq\Lambda) - t_{11}$ around $(q_0, \omega_0)$ and keep to the lowest order of $\delta$, we have

$$t_{12} = \exp(-ik_a d_a) \left[ \frac{i}{2} \left( \frac{Z_a}{Z_b} - \frac{Z_b}{Z_a} \right) \right] \left[ (-1)^m \delta + O^3(\delta) \right], \quad \text{(B9)}$$

$$\exp(iq\Lambda) - t_{11} \propto O^2(\delta). \quad \text{(B10)}$$

Since electric field is linear combination of $t_{12}$ and $\exp(iq\Lambda) - t_{11}$, it will change sign near $(q_0, \omega_0)$, i.e., $|u_{nq_0^+}\rangle = -|u_{nq_0^-}\rangle$, where $|u_{n,q}\rangle$ is the normalized periodic part of the field in cell eigenvector at $(\omega, q)$ of the $n^{th}$ band and $X^{+(-)}$ means approaching $X$ from the positive (negative) direction.

At other band point except $(q_0, \omega_0)$, $|u_{n,q}\rangle$ is a continuous function of $q$. Since the inversion center is chosen at the origin, the system is invariant under the space inversion. Following the same argument as stated before, we can find

$$E_{n,q}(z) = E_{n,-q}(-z) \quad \text{(B11)}$$

With $E_{x;n,q}(z) = u_{n,q}(z) \exp(iqz)$, we have $u_{n,q}(z) = u_{n,-q}(-z)$. Since $|u_{n,q}\rangle$ is periodic part of the wave function, the integration in $\text{Im}\langle u_{n,q} | \partial_q | u_{n,q} \rangle$ is performed from $-\Lambda/2$ to $\Lambda/2$. Thus $\text{Im}\langle u_{n,q} | \partial_q | u_{n,q} \rangle$ is an odd function of $q$, in other words

$$\text{Im}\left[ \langle u_{n,q} | \partial_q | u_{n,q} \rangle + \langle u_{n,-q} | \partial_q | u_{n,-q} \rangle \right] = 0. \quad \text{(B12)}$$

If $t_{12}$ and $\exp(iq\Lambda) - t_{11}$ do not equal to 0 simultaneously on one band, then with the chosen gauge, $|u_{n,q}\rangle$ is a continuous function of $q$. Thus (B12) is applied all over that band and the Zak phase is 0. Otherwise, if $t_{12}$ and $\exp(iq\Lambda) - t_{11}$ equal to 0 simultaneously at $(q_0, \omega_0)$, then $|u_{n,q}\rangle$ is discontinuous at $(q_0, \omega_0)$, and the Zak phase of this band is given by

$$\gamma_n = -\operatorname{Im} \lim_{\delta q \to 0} \left\{ \left( \int_{-\pi/\Lambda}^{q_0-\delta q} + \int_{q_0+\delta q}^{\pi/\Lambda} \right) dq \langle u_{n,q} | \partial_q | u_{n,q} \rangle + \ln \langle u_{n,q_0+\delta q} | u_{n,q_0-\delta q} \rangle \right\} = \pi. \quad (B13)$$

The above proof can be easily extended to the case when the system is dispersive [12-14].

# APPENDIX C: EIGEN STATE AT THE BAND EDGE

In this appendix, we will prove that, the electric field at the inversion center for band edge state should be zero or maximum. There are two inversion centers in this system, namely the center of A slab and B slab. Without loss of generality, we choose the center of A slab as the inversion center. At the center of A slab, according to Eqs. (B3) and (B4), $E_x = \zeta_E \{ t_{12} \exp(ik_a d_a) + [\exp(iq\Lambda) - t_{11}] \}$, $H_y = \zeta_H \{ t_{12} \exp(ik_a d_a) - [\exp(iq\Lambda) - t_{11}] \}$, where $\zeta_E$ and $\zeta_H$ are some complex constants. For an arbitrary state $(\omega, q)$ on the band, $t_{12} \exp(ik_a d_a)$ and $\exp(iq\Lambda) - t_{11}$ are pure imaginary. At band edges, $\cos(q\Lambda) = \pm 1$, $\sin(q\Lambda) = 0$. After some mathematics, we could arrive at

$$\{\text{RHS of Eq.(A1)}\}^2 + \{\operatorname{Im}[\exp(iq\Lambda) - t_{11}]\}^2 = 1 + \{\operatorname{Im}[t_{12} \exp(ik_a d_a)]\}^2, \quad (C1)$$

then

$$t_{12} \exp(ik_a d_a) = \pm [\exp(iq\Lambda) - t_{11}]. \quad (C2)$$

From Eq. (C2), it is easy to find out that either the electric field or magnetic field should be 0 at the center of A slab. So there are only two types of states at the band edges. For type A (Anti-symmetry) state, as $E_x(z=0) = 0$. For type S (Symmetry) state, as $E_x(z=0) \neq 0$, $H_y(z=0) = 0$; electric field is at the maximum value inside A slab.

Now we go further to find out whether A or S state is at the lower or upper edge of the $n^{th}$ gap. The sign of the function $\sin(\omega n_b d_b / c)$ depends on the number of zeros it crosses in the frequency range $(0, \omega)$, so to get $\operatorname{sgn}(\sin k_b d_b)$, we only need to count the number of zero points of $\sin k_b d_b$. As proved before, if the frequency $\tilde{\omega}$ at which $\sin k_b d_b = 0$ is on an isolated band, then this band has Zak phase $\pi$, otherwise, $\tilde{\omega}$ is at the crossing point of two bands since $\tilde{\omega}$ is always on pass band. Thus, for a frequency $\omega$ inside the n$^{th}$ gap,

$$\text{sgn}\left(\sin\frac{\omega}{c}n_b d_b\right) = (-1)^l \exp\left(i\sum_{m=1}^{n-1}\theta_m^{Zak}\right), \quad \text{(C3)}$$

where $l \in \mathbb{N}$ is the number of band crossing points under the $n^{th}$ gap, and the second term on the RHS of Eq. (C3) is a summation of Zak phase below this gap. Define $\chi = \text{sgn}(1-\varepsilon_a\mu_b/\varepsilon_b\mu_a)$, then according to Eq. (B2),

$$\text{sgn}\{\text{Im}[t_{12}\exp(ik_a d_a)]\} = (-1)^l \exp\left(i\sum_{m=1}^{n-1}\theta_m^{Zak}\right)\chi. \quad \text{(C4)}$$

With Eq. (B1), it is easy to get, for band edge states

$$\text{Im}[\exp(iq\Lambda) - t_{11}] = -[\sin(\gamma) + (\tau-1)\cos k_a d_a \sin k_b d_b]. \quad \text{(C5)}$$

At $\gamma = (2n+1/2)\pi$, where $n \in \mathbb{N}$, $\sin(\gamma) = 1$, $\cos k_a d_a \sin k_b d_b = \cos^2(k_a d_a) \geq 0$; at $\gamma = (2n+3/2)\pi$, where $n \in \mathbb{N}$, $\sin(\gamma) = -1$, $\cos k_a d_a \sin k_b d_b = -\cos^2(k_a d_a) \leq 0$. Thus at $\gamma = (n+1/2)\pi$, where $n \in \mathbb{N}$,

$$\text{sgn}[\sin(\gamma) + (\tau-1)\cos k_a d_a \sin k_b d_b] = \text{sgn}[\sin(\gamma)]. \quad \text{(C6)}$$

From Eq. (C1), it is easy to get that, for states on the band,

$$[\sin(\gamma) + (\tau-1)\cos k_a d_a \sin k_b d_b]^2 = (\tau^2-1)\sin^2 k_b d_b + 1 - \cos(q\Lambda) \geq 0. \quad \text{(C7)}$$

The equality is achieved at the point where two bands cross, thus $[\sin(\gamma) + (\tau-1)\cos k_a d_a \sin k_b d_b]$ does not change sign on the isolated pass band. $\sin(\gamma)$ only changes sign inside band gap (or at the crossing point of two band) and the frequency at which $\gamma = (n+1/2)\pi$ must be in the pass band, so Eq. (C6) is also true for band edge states. As $\gamma = n\pi$ ($n \in \mathbb{N}^+$) gives the mid-gap position of the $n^{th}$ gap, for the edge state below the $n^{th}$ gap,

$$\text{sgn}\{\text{Im}[\exp(iq\Lambda)-t_{11}]\} = -\text{sgn}[\sin(\gamma)] = (-1)^n, \quad \text{for the edge state above the } n^{th} \text{ gap,}$$

$$\text{sgn}\{\text{Im}[\exp(iq\Lambda)-t_{11}]\} = (-1)^{n+1}.$$

In all, if $(-1)^n(-1)^l \exp\left(i\sum_{m=1}^{n-1}\theta_m^{Zak}\right)\chi = 1$, then the edge state below the $n^{th}$ gap is a S state, above the $n^{th}$ gap is A state; otherwise if $(-1)^n(-1)^l \exp\left(i\sum_{m=1}^{n-1}\theta_m^{Zak}\right)\chi = -1$, then the state below the $n^{th}$ gap is A state, above the $n^{th}$ gap is S state.

In Fig. 5, we give an example to illustrate the relation between edge state and Zak phase. The band structure (solid black line) of a particular PC with parameters given by $\varepsilon_a = 4$, $\mu_a = \varepsilon_b = \mu_b = 1$, $d_a = 0.4\Lambda$ and $d_b = 0.6\Lambda$ is plotted in Fig. 5. The rule specified in Appendix B gives the Zak phase of each isolated band, as showed with green letter. We labeled the type A edge states with solid purple circle and S state with yellow circle. Same as Fig. 1(b) in the main text, $\text{sgn}[\varsigma] = \text{sgn}[\text{Im}(Z_s)]$ is marked by magenta color when $\varsigma > 0$ and cyan when $\varsigma < 0$. Three important features should be pointed out in Fig. 5: (i) The state must change from (S) to (A) or from (A) to (S) when passing through a band gap; (II) The upper and lower edge states of a band are of the same type if the Zak phase of this band is 0, otherwise it is $\pi$; (III) $\text{sgn}[\varsigma]$ of each gap is related to the edge state bounding this gap. If (A) state is at the lower edge and (S) state is at the upper edge, then $\varsigma < 0$, otherwise, if (S) state is at the lower edge and (A) states is at the upper edge, $\varsigma > 0$. This will be proved in the next section.

## APPENDIX D: BULK BAND AND SIGN OF REFLECTION PHASE

In this appendix, we will show that $\text{sgn}(\phi_n) = (-1)^n(-1)^l \exp(i\sum_{m=1}^{n-1}\theta_m^{Zak})\chi$, where $\phi_n$ is the reflection phase of the $n^{th}$ gap (as defined below).

We consider a plane wave $E_i = E_0 e^{ikz}$ being incident on the PC from vacuum as shown in Fig. 1(a) in main text, and the reflected wave is $E_r = rE_0 e^{-ikz}$. The field inside the gap at $z = 0^+$ is given by

$$E_x = t_{12} \exp(ik_a d_a / 2) + \left[\exp(iq\Lambda) - t_{11}\right] \exp(-ik_a d_a / 2), \quad (D1)$$

$$H_y = \frac{k_a}{\omega \mu_a} \left\{ t_{12} \exp(ik_a d_a / 2) - \left[\exp(iq\Lambda) - t_{11}\right] \exp(-ik_a d_a / 2) \right\}. \quad (D2)$$

Matching the boundary condition, we have

$$\frac{1+r}{1-r} = z_a \frac{t_{12} \exp(ik_a d_a) + \left[\exp(iq\Lambda) - t_{11}\right]}{t_{12} \exp(ik_a d_a) - \left[\exp(iq\Lambda) - t_{11}\right]}, \quad (D3)$$

From Eq. (D3) we could calculate the reflection phase delay inside the gap. As an example, in Fig. 6, we give the reflection phase delay inside the first gap of "PC3" ($\varepsilon_b = 3.5$, $\varepsilon_a = \mu_a = \mu_b = 1$, $d_a = 0.35\Lambda$ and $d_b = 0.65\Lambda$) for light incident from vacuum (black line). The reflection phase increases monotonically from $-\pi$ to $0$ with increasing frequency. As a comparison, we also give the reflection phase calculated directly using transfer-matrix for a slab consisting of a finite number of unit cells of PC3. The solid blue line is for a slab with 5 unit cells and solid red line is for a slab with 10 unit cells of PC3. As the number of unit cell increases, the reflection phase converges to the one given by Eq. (D3), which is derived for a semi-infinite PC.

For frequency inside the $n^{th}$ gap (including band edge states), $q\Lambda = n\pi + i\mathrm{K}$ [18], where $\mathrm{K} > 0$ and describes the decay length inside the gap.

$$\left| t_{12} \exp(ik_a d_a) \right|^2 - \left| \mathrm{Im}\left[ \exp(iq\Lambda) - t_{11} \right] \right|^2 = \cosh^2(\mathrm{K}) - 1 \geq 0, \quad (D4)$$

and the equality is only achieved at the band edge. From Eq. (B2), we know $t_{12} \exp(ik_a d_a)$ is a pure imaginary number, so

$$\text{sgn Im}\{t_{12}\exp(ik_a d_a)\pm[\exp(iq\Lambda)-t_{11}]\}=\text{sgn Im}[t_{12}\exp(ik_a d_a)]=(-1)^l \exp\left(i\sum_{m=1}^{n-1}\theta_m^{Zak}\right)\chi \quad .(D5)$$

Inside the $n^{th}$ gap,

$$\text{Re}\left[\exp(iq\Lambda)-t_{11}\right]=(-1)^{n+1}\sinh K \, . \text{(D6)}$$

So

$$\frac{1+r}{1-r}=z_a\frac{(-1)^n(-1)^l \exp\left(i\sum_{m=1}^{n-1}\theta_m^{Zak}\right)\chi\lambda_+ + i}{(-1)^n(-1)^l \exp\left(i\sum_{m=1}^{n-1}\theta_m^{Zak}\right)\chi\lambda_- - i}, \text{(D7)}$$

where $\lambda_\pm = \left|t_{12}\exp(ik_a d_a)\pm i\,\text{Im}[\exp(iq\Lambda)-t_{11}]\right|/\sinh K > 0$. Inside the $n^{th}$ gap, $r = e^{i\phi_n}$, where $\phi_n$ (the subscript n labels the gap) is a function of frequency. Thus $\frac{1+r}{1-r}=i\cot(\phi_n/2)$, a pure imaginary number. So the RHS of Eq. (D7) is also pure imaginary, then $\lambda_+\lambda_- = 1$

$$\text{Im}\left[z_a\frac{(-1)^n(-1)^l \exp\left(i\sum_{m=1}^{n-1}\theta_m^{Zak}\right)\chi\lambda_+ + i}{(-1)^n(-1)^l \exp\left(i\sum_{m=1}^{n-1}\theta_m^{Zak}\right)\chi\lambda_- - i}\right]=\text{sgn}\left((-1)^n(-1)^l \exp\left(i\sum_{m=1}^{n-1}\theta_m^{Zak}\right)\chi\right)z_a\lambda_+ . \text{(D8)}$$

So if the reflection phase is limited to $[-\pi,\pi]$ and from the relation $\phi_n = \pi - 2\arctan(\varsigma_n)$ (here we shift $\phi_n$ back to be inside $[-\pi,\pi]$) then

$$\text{sgn}[\varsigma_n]=\text{sgn}(\phi_n)=(-1)^n(-1)^l \exp(i\sum_{m=1}^{n-1}\theta_m^{Zak})\chi \, . \quad (D9)$$

It is easy to show that $\text{Im}[\exp(iq\Lambda)-t_{11}]$ is a monotonic function inside the $n^{th}$ gap. As $\lambda_+\lambda_- = 1$, and the different between $\lambda_+, \lambda_-$ is $\text{Im}[\exp(iq\Lambda)-t_{11}]$, thus $\lambda_+$ is a monotonic function inside the $n^{th}$ gap. Combining Eqs. (D7) and (D8), we find $\phi_n$ is also a monotonic function. As we have already proved, the edge state must be A state or S state. When A is present, $r = -1$ and $\phi = \pm\pi$; when S is presented, $r = 1$ and $\phi = 0$. With Eq. (D8) and the reflection phase at band edge, we could further conclude that $\phi_n$ is a monotonic increasing function of frequency. From the relation $\phi_n = \pi - 2\arctan(\varsigma_n)$, it is straight forward to show that, $\varsigma_n$ is a monotonic decreasing function of frequency from $\infty$ to 0 or from 0 to $-\infty$ depending on the sign of $\varsigma_n$.

In the main text, we argue that, once $\text{sgn}\,\varsigma$ of the left and right PCs are different inside the common band gap, there must be an interface state, here we will give another example. In Fig. 7(b), we choose the 2$^{\text{nd}}$ common band gap in Fig. 4 of the main text as an example. The solid black, red and blue lines show the imaginary parts of relative impedances of PC3, PC4 and the sum of those two, respectively, inside the 2$^{\text{nd}}$ common band gap. The solid black and red lines are both monotonic decreasing function of frequency, and their sum must also be a monotonic decreasing function of frequency from positive to negative. Thus there must exist some frequency point at which the blue line crosses the 0, corresponding to an interface state as showed in Fig. 7(a), where resonant transmission is observed inside the common band gap.

# APPENDIX E: EXTENTION OF EQ. (4)

In this appendix, we will show that, Eq. (4) in the main text is still valid when the relative permittivity and permeability are continuously varying functions, and the lattice constants of the left and right periodic systems do not need to be equal.

For an interface state to exist, we only need two overlapped gaps with different signs of $\varsigma = \text{Im}[Z_S / Z_0]$ and we do not care about the "origin" of the gap (e.g. gap number or the lattice constants of PCs of the left or right periodic system). In Fig. 8 we give an example to illustrate this point. We consider a system consisting of 7 unit cells of "PC5" ( $\varepsilon_b = 3.5$, $\varepsilon_a = \mu_a = \mu_b = 1$, $d_a = 0.7\Lambda$ and $d_b = 1.3\Lambda$ ) on the left and 14 unit cells of "PC4" on the right embedded in vacuum. The parameters of PC4 are same as that in the main text, i.e., $\mu_b = 6$, $\varepsilon_a = \mu_a = \varepsilon_b = 1$, $d_a = 0.6\Lambda$ and $d_b = 0.4\Lambda$. Note here that we just double the length of $d_a$ and $d_b$ of "PC3" in the main text, so that the lattice constant of PC5 is twice that of PC4. According to the scaling law, the Zak phase of each isolated band would not change, as labeled in Fig. 8 (b) and (c) with green letter. Using Eq.(4) in main text, we could get the $\text{sgn}[\varsigma]$, which is also labeled with magenta ( $\varsigma > 0$ ) or cyan color ( $\varsigma < 0$ ) in Fig. 8 (b) and (c). Though now the gap number of PCs from the left and right side of common gap region are different, the rule still applies as whenever two gaps with different color have common frequency region, there must be an interface state.

Eq. (4) in the main text also applies when the dielectric function is a continuous function of z. In Fig. 9, we considered a system consist of 20 unit cells of "PC6" ($\varepsilon_r = 12+6\sin[2\pi(z/\Lambda+1/4)]$, $\mu_r = 1$) on the left side and 10 unit cells of "PC7" ($\varepsilon_r = 12+5\sin[2\pi(z/\Lambda-1/4)]+5\sin[4\pi(z/\Lambda+1/8)]$, $\mu_r = 1$) on the right hand side embedded in vacuum. The transmission spectrum of system is given in Fig. 9 (a), where the boundary between two PCs is set at $z = 0$. The band structures (Solid black line) of PC6 and PC7 are given in Fig. 9 (b) and (c), respectively. We calculated the Zak phase of each band numerically with Eq. (3) and labeled with green letter, then the $\text{sgn}[\varsigma]$ of each gap is also shown with magenta ($\varsigma > 0$) or cyan color ($\varsigma < 0$) in Fig. 9 (b) and (c). It is clear that, Eq. (4) could still predict the existence or absence of the interface state in this case.

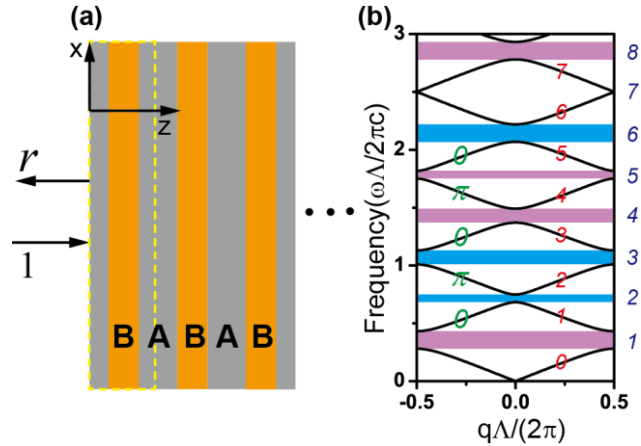

FIG. 1. (a), Plane wave incidents normally on an AB layered structure, the reflection coefficient of electric field is given by $r$. The yellow dash line marks the unit cell we considered. (b), The band structure of PC (solid black curve) with parameters given by $\varepsilon_a = 4$, $\mu_a = \varepsilon_b = \mu_b = 1$, $d_a = 0.4\Lambda$ and $d_b = 0.6\Lambda$ and $\Lambda$ is the length of unit cell. The magenta strip means gap with $\varsigma > 0$, while cyan strip means gap with $\varsigma < 0$, the Zak phase of each individual

band is labeled with green letter, and the number of bands and gaps are listed with red and blue letter, respectively.

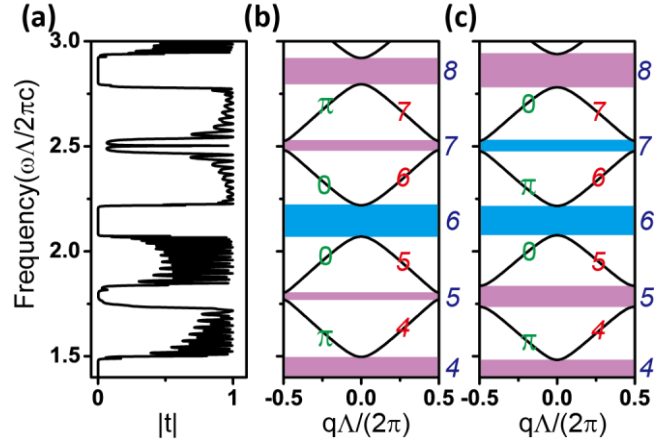

FIG. 2. (a), The transmission spectrum of system consisted with 10 unit of PC1 on the left side and 10 unit of PC2 on the right side in vacuum, Where the parameters of PC1 are given by $\varepsilon_a = 3.8$, $\varepsilon_b = \mu_a = \mu_b = 1$, $d_a = 0.42\Lambda$ and $d_b = 0.58\Lambda$, the parameters of PC2 are given by $\varepsilon_a = 4.2$, $\varepsilon_b = \mu_a = \varepsilon_b = 1$, $d_a = 0.38\Lambda$ and $d_b = 0.62\Lambda$, and $\Lambda$ is the unit length of PCs. (b), (c), The band structure (solid black curve) of PC1 and PC2. In both (b) and (c) the magenta strip means gap with $\varsigma > 0$, while cyan strip means gap with $\varsigma < 0$, and the Zak phase of each individual band is also labeled in (b) and (c) with green letter. We note that if the gaps of the PCs on either side of the interface carry the same sign of $\varsigma$, there is no interface state. If the sign of $\varsigma$ is opposite, there must be an interface state (e.g. at reduced frequency unit of 2.5).

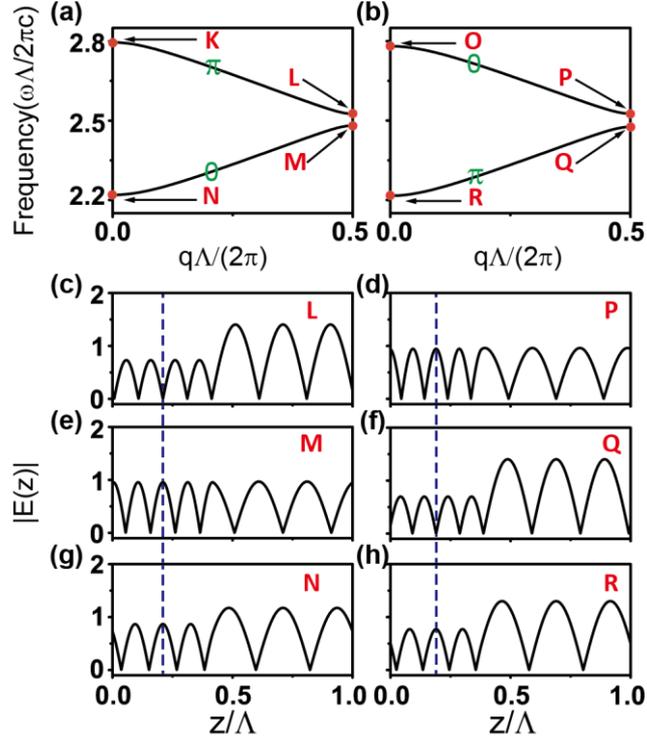

FIG. 3. (a), The band dispersion of the PC with parameters $\varepsilon_a = 3.8$, $\varepsilon_b = \mu_a = \mu_b = 1$, $d_a = 0.42\Lambda$ and $d_b = 0.58\Lambda$. (b), The band dispersion of the PC with parameters $\varepsilon_a = 4.2$, $\varepsilon_b = \mu_a = \mu_b = 1$, $d_a = 0.38\Lambda$ and $d_b = 0.62\Lambda$ and $\Lambda$ is the unit length of PCs. The Zak phase of each band in (a) and (b) are shown with green letters. (c-h), The absolute value of electric field $E(z)$ (black solid line) of band edge state as a function of position z, six band edge states (L, P, M, Q, N, R, indicated with solid red circle in (a) and (b)) are shown in (c), (d), (e), (f), (g), (h), respectively. The region of A slab is $(0, d_a)$, the left is B slab, and the blue dash lines mark the center of A slab.

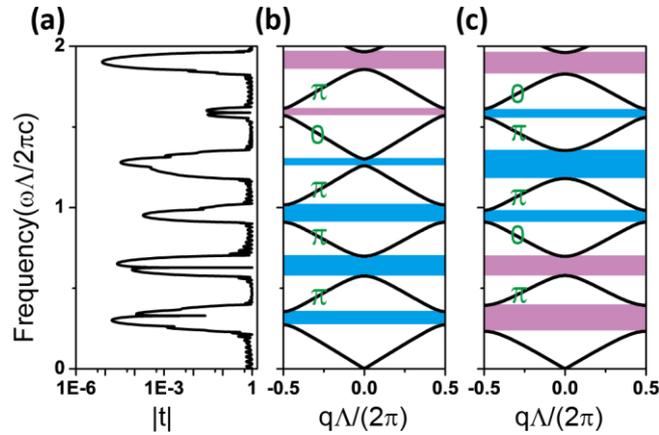

FIG. 4. (a), The transmission spectrum of system consisted of 10 unit of PC3 on the left side and 10 unit of PC4 on the right hand side in vacuum. Where the parameters of PC3 are given by $\varepsilon_b = 3.5$, $\varepsilon_a = \mu_a = \mu_b = 1$, $d_a = 0.35\Lambda$ and $d_b = 0.65\Lambda$, the parameters of PC4 are given by $\mu_b = 6$, $\varepsilon_a = \mu_a = \varepsilon_b = 1$, $d_a = 0.6\Lambda$ and $d_b = 0.4\Lambda$, and $\Lambda$ is the unit length of PCs. (b), (c), The band structure (solid black curve) of PC3 and PC4. In both (b) and (c) the magenta strip means gap with $\varsigma > 0$, while cyan strip means gap with $\varsigma < 0$, and the Zak phase of each individual band is also labeled in (b) and (c) with green letter. Whenever two gaps with different character (different sign of $\varsigma$) have a common region, there will be an interface state.

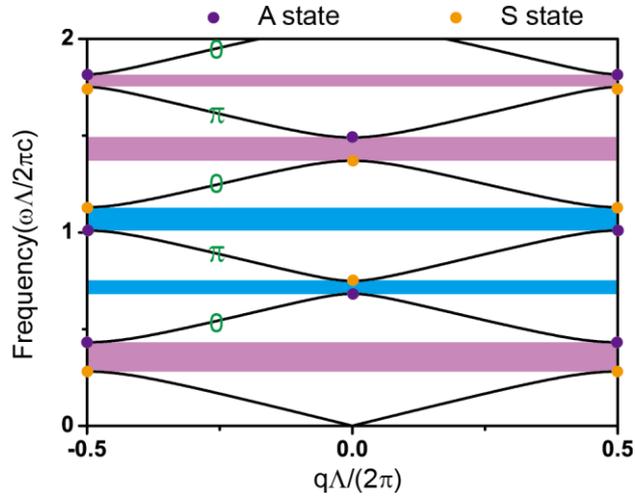

FIG. 5. The band structure (solid black line) of PC with parameters given by $\varepsilon_a = 4$, $\mu_a = \varepsilon_b = \mu_b = 1$, $d_a = 0.4\Lambda$ and $d_b = 0.6\Lambda$. The light magenta strip means gap with $\varsigma > 0$, while cyan strip means gap with $\varsigma < 0$, the Zak phase of each individual band is labeled with green letter. The solid purple circle means A (Anti-symmetric) state ($E_{eig} = 0$ at the center of A slab) at the band edge and the solid yellow circle means S (Symmetric) state ($E_{eig} \neq 0$ at the center of A slab).

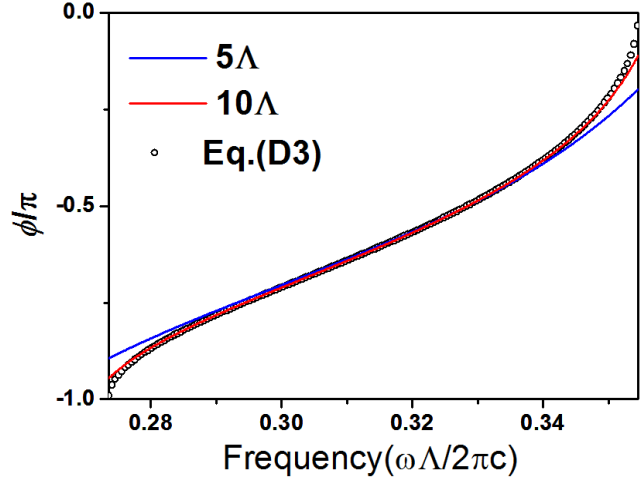

FIG. 6. The reflection phase of PC3 in vacuum inside the first band gap. The parameters of PC3 are given by $\varepsilon_b = 3.5$, $\varepsilon_a = \mu_a = \mu_b = 1$, $d_a = 0.35\Lambda$ and $d_b = 0.65\Lambda$. The solid blue line and solid red line are calculated with transfer matrix directly and for 5, 10 unit cells of PC3, respectively. The open black circle is calculated with the Eq. (D3). It is clearly that, as the unit number increase, the reflection phase converges to the one calculated with the Eq. (D3).

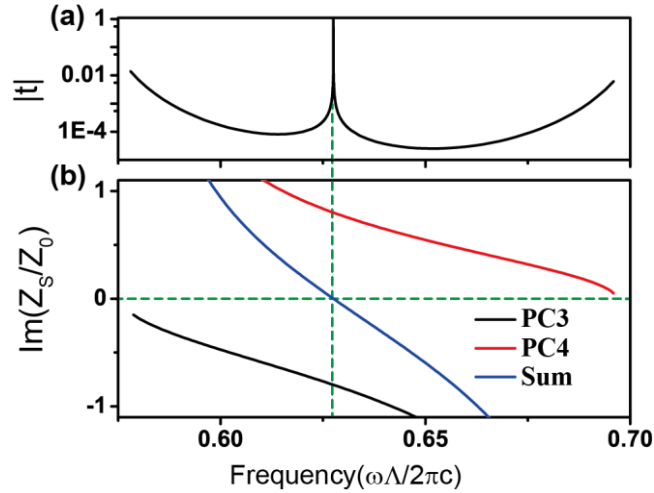

FIG. 7. (a), The transmission spectrum of system consisted with 10 unit of PC3 on the left side and 10 unit of PC4 on the right side in vacuum, Where the parameters of PC3 are given by $\varepsilon_b = 3.5$, $\varepsilon_a = \mu_a = \mu_b = 1$, $d_a = 0.35\Lambda$ and $d_b = 0.65\Lambda$, the parameters of PC4 are given by $\mu_b = 6$, $\varepsilon_a = \mu_a = \varepsilon_b = 1$, $d_a = 0.6\Lambda$ and $d_b = 0.4\Lambda$, and $\Lambda$ is the unit length of PCs. (b), The imaginary part of relative surface impedance (divided by the impedance of vacuum) of PC3 (solid black), PC4 (solid red) and summation of those two (solid blue) inside the common gap region. The green dash lines are just drew for illustration, the position where blue line cross 0 corresponds to an interface state between two PCs.

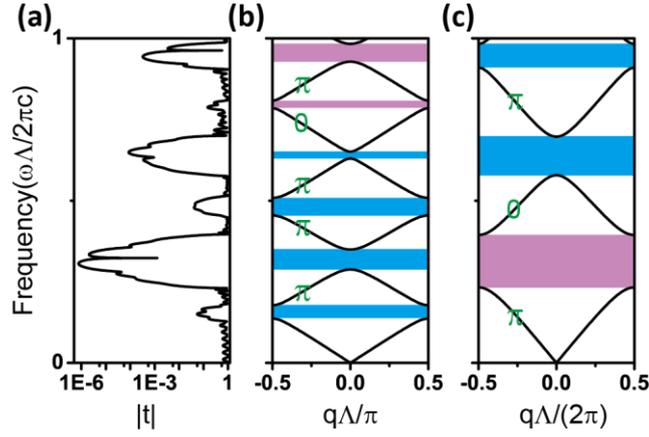

FIG. 8. (a), The transmission spectrum of system constructed with 7 unit cells of PC5 on the left side and 14 unit cell of PC4 on the right hand side in vacuum. Where the parameters of PC5 are given by $\varepsilon_b = 3.5$, $\varepsilon_a = \mu_a = \mu_b = 1$, $d_a = 0.7\Lambda$ and $d_b = 1.3\Lambda$, which just doubles the length of PC3, the parameters of PC4 are given by $\mu_b = 6$, $\varepsilon_a = \mu_a = \varepsilon_b = 1$, $d_a = 0.6\Lambda$ and $d_b = 0.4\Lambda$. (b), (c), The band structure (solid black curve) of PC5 and PC4. In both b and c, the magenta strip means gap with $\varsigma > 0$, while cyan strip means gap with $\varsigma < 0$, and the Zak phase of each individual band is also labeled in (b) and (c) with green letter.

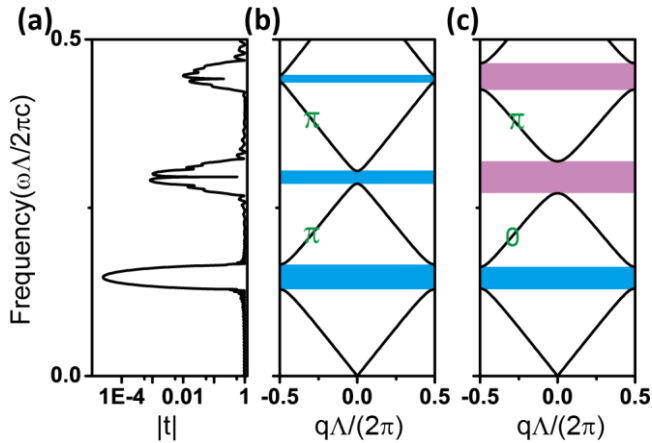

FIG. 9. (a), The transmission spectrum of system constructed with 20 unit of PC6 on the left side and 10 unit of PC7 on the right hand side in vacuum, where the parameters of PC6 are given by $\varepsilon_r = 12+6\sin[2\pi(z/\Lambda+1/4)]$, $\mu_r = 1$, the parameters of PC7 are given by $\varepsilon_r = 12+5\sin[2\pi(z/\Lambda-1/4)]+5\sin[4\pi(z/\Lambda+1/8)]$, $\mu_r = 1$. The boundary between PC6 and PC7 now is set at z=0. (b), (c), The band structure (solid black line) of PC6 and PC7. In both (b) and (c), the magenta strip means gap with $\varsigma > 0$, while cyan strip means gap with $\varsigma < 0$, and the Zak phase of each individual band is also labeled in (b) and (c) with green letter.